\documentclass[%
 reprint,
 amsmath,amssymb,
 aps,
]{revtex4-2}

\usepackage{graphicx}
\usepackage{dcolumn}
\usepackage{bm}

\begin{document}

\preprint{APS/123-QED}

\title{Microwave-induced inverse Faraday effect in superconductors}

\author{A. Hamed Majedi}
 \affiliation{Department of Electrical and Computer Engineering, Department of Physics and Astronomy\\ Waterloo Institute for Nanotechnology, University of Waterloo, Waterloo, Ontario, Canada N2L 3G1} 
 \altaffiliation[Also at ]{Perimeter Institute for Theoretical Physics, Waterloo, Ontario, Canada.}

\date{\today}

\begin{abstract}
Inverse Faraday effect (IFE) in superconductors is proposed, where a static magnetization is generated under the influence of a circularly polarized microwave field. Classical modeling of the IFE explicitly provides superconducting gyration coefficient in terms of its complex conductivity. IFE is then considered as a source of nonlinearity and gyrotropy even at a low-power microwave regime giving rise to a spectrum of phenomena and applications. Microwave-induced gyroelectric conductivity, Hall effect, microwave birefringence, flux quantization and vortex state are predicted and quantitatively analyzed. Peculiar microwave birefringence in gyrotropic superconductors due to radical response of superelectrons has been highlighted.     
\end{abstract}

\maketitle

\noindent
{\it Introduction}-
Nonlinear microwave response of superconducting structures is a core subject not only in probing the physics of superconductivity \cite{lara2015microwave} but also implemented in numerous applications ranging from quantum metrology to superconducting qubits and microwave quantum optics \cite{you2011atomic}. The microwave nonlinearities in superconducting structure possess diverse origins but mostly are attributed to nonlinear kinetic \cite{anlage1989current,mohebbi2009periodic} and Josephson inductance \cite{rossignol2019role,mohebbi2009analysis} including Kerr-type \cite{robson2017giant,krupko2018kerr}, Duffing and anharmonicity \cite{manucharyan2007microwave,nigg2012black}, weak links \cite{kummel1990andreev,abdo2006observation,ghamsari2009current}, phase slip formation \cite{jafari2016stimulated} and vortex dynamics \cite{gurevich1999nonlinear, yip1992nonlinear}.\\
Borrowing from optomagnetics, a less-explored field in nonlinear optics \cite{majedi2020nonlinear}, a microwave-induced nonlinearity in superconductors based on the inverse Faraday effect (IFE) is proposed in this letter. IFE refers to the generation of static magnetic field by not linearly polarized, e.g. circularly polarized, light \cite{pitaevskii1961electric}. Purely nonlinear effect arising from IFE is solely based on the gyration of the time-varying electric field and it does not directly link to any linear electromagnetic properties of the materials such as Kerr-type that is related to the linear refractive index. IFE in superconductor is based on angular momentum transfer between the circularly polarized microwave field and superconductor that is considered as superelectron condensate and normal electrons. Electric field gyration creates encircling supercurrent and normal current associated with local static magnetic field opposing the microwave field. IFE can be employed not only to make tunable gyrotropic superconductor prevailing upon phenomena such as Hall effect and microwave birefringence but also to derive type II superconductors to vortex state.\\
In this letter I first develop a classical formalism to find the microwave induced static magnetic field through a gyration coefficient that is proportional to complex conductivity of superconductor. The critical microwave field to suppress 
 superconductivity is derived in terms of the critical magnetic field. Turning to IFE consequences in superconductors, the gyroelectric conductivity and microwave-induced Hall effect, as an experimental tool to measure superconducting gyration coefficient are analytically discussed. Gyroelectric conductivity is also used for linear and circular birefringence for a microwave pump-probe scenario. Finally, I propose an embodiment of the dynamic and controllable flux quantization and vortex state in type II superconductors using circularly polarized microwave field.\\                           
{\it Modeling of Inverse Faraday Effect in Superconductors}-
The electrodynamic response of a superconductor is considered in the Two-Fluid model. The equation of motion for superelectrons and normal electrons under the influence of electric field ${\bf E}({\bf r},t)$ and its associated magnetic field ${\bf B}({\bf r},t)$ can be phenomenologically described by the London equations
\begin{eqnarray}
\label{2F1}
m\frac{d}{dt}{\bf v}_s&=&e{\bf E}({\bf r},t)+e{\bf v}_s\times{\bf B}({\bf r},t)\\
\label{2F2}
m\frac{d}{dt}\langle {\bf v}_n\rangle+m\Gamma \langle {\bf v}_n\rangle&=&e{\bf E}({\bf r},t)+e\langle {\bf v}_n\rangle\times{\bf B}({\bf r},t)
\end{eqnarray}
where $m$ is the mass of an electron, $e$ is charge of an electron, $\Gamma=\displaystyle{\frac{1}{\tau}}$ is the inverse of momentum relaxation time for normal electrons and ${\bf v}_s$ and $\langle {\bf v}_n\rangle$ are the superelectron and average normal electron velocities, respectively. We consider a circularly polarized plane electromagnetic field with pumping frequency, $\omega_p$, traveling normal to the surface of a semi-infinite superconductor, i.e. at $z=0$, in the following form
\begin{eqnarray}
\label{Efield}
&{\bf E}&=\mbox{Re}\{{\bf{\tilde E}}e^{i\omega_p t}\}=\mbox{Re}\Big\{E_o({\bf x}+i{\bf y})e^{-\alpha_p z}e^{i(\omega_p t-\beta_p z)}\Big\} \ \ \\
\label{Bfield}    
&{\bf B}&=\mbox{Re}\{{\bf{\tilde B}}e^{i\omega_p t}\}=\mbox{Re}\Big\{B_o({\bf y}-i{\bf x})e^{-\alpha_p z}e^{i(\omega_p t-\beta_p z)}\Big\} \ \ 
\end{eqnarray}
where ${\bf x},{\bf y},{\bf z}$ are Cartesian unit vectors,  $0<\hbar\omega_p<2\Delta$, $\Delta$ being a superconducting energy gap, and $\alpha_p$ and $\beta_p$ are the propagation loss and the propagation constant, respectively. The low-frequency propagation characteristics can be derived based on the Two-Fluid model \cite{lancaster2006passive}. When the circularly polarized wave interacts with the ensemble of free superelectrons and normal electrons, its angular momentum generates local circulating supercurrent and normal current in the superconductor that produces a magnetic field parallel to the microwave field opposing in the direction. This is a manifestation of the IFE where high-intensity circularly polarized light generates DC magnetization in matter, a theoretical prediction by L. Pitaevskii in 1961 \cite{pitaevskii1961electric} and its subsequent experimental demonstration in 1965 \cite{van1965optically}.    
The magnetization in superconductor, ${\bf M}_s$ can be found as \cite{majedi2020nonlinear}
\begin{equation}
\begin{aligned}
    {\bf M}_s=
    \frac{n_s e}{2m}{\bf L}_s+\frac{n_n e}{2m}{\bf L}_n 
\end{aligned}    
\end{equation}
where ${\bf L}_s={\bf r}_s\times{\bf p}_s$ is the magnetic moment of superelectrons and ${\bf L}_n={\bf r}_n\times\langle {\bf p}_n\rangle$ is the magnetic moment of normal electrons in terms of their momenta ${\bf p}_s$ and $\langle {\bf p}_n\rangle$, respectively. For intermediate temperature range, $0<T<T_c$, both superelectrons with density number $n_s(T)$ and normal electrons with density number $n_n(T)$ coexist in the form of
\begin{equation}
   n= n_s(T)+n_n(T)=n\big(1-(\frac{T}{T_c})^s\big)+n\big(\frac{T}{T_c}\big)^s
\end{equation}
where $s$ is an empirical exponent, i.e. $s=4$ for low-temperature superconductors and $s=2$ for high-temperature superconductors and $n$ is the total number density \cite{vendik1998empirical}.
Considering the microwave signal given in Eqs. (\ref{Efield}) and (\ref{Bfield}), the solutions of Eqs. (\ref{2F1}) and (\ref{2F2}) can yield the DC magnetization in the superconductor as
\begin{equation}
\label{Pita1}
{\bf M}_{DC}=i\gamma(\omega_p,T){\bf{\tilde E}}\times{\bf{\tilde E}}^*
\end{equation}
where the gyration coefficient $\gamma(\omega_p,T)$ is
\begin{eqnarray}
\label{Pita2}
\nonumber
\gamma(\omega_p,T)&=&
\frac{-e^3}{4m^2\omega_p}\Big(\frac{n_s}{\omega^2_p}+\frac{n_n}{\omega^2_p+\Gamma^2}\Big)\\
&\approx&
\frac{-e}{4m\omega_p}\bigg(\frac{\sigma_2}{\omega_p}+\frac{\sigma_1}{\Gamma}\bigg)
\end{eqnarray}
Eq. (\ref{Pita1}) is in the form of Pitaevskii's relationship \cite{pitaevskii1961electric} indicating that the DC magnetization is solely depends on the gyration of the electric field, through the relationship of $i{\bf\tilde E}\times{\bf \tilde E}^*=2{\bf z}|E_o|^2e^{-2\alpha_p z}$. This fact is signified by the gyration coefficient, $\gamma(\omega_p,T)$, reminiscent to magnetogyration coefficients in magneto-optic materials \cite{saleh2019fundamentals}.
The microwave-induced DC magnetization in superconductor is inherently a nonlinear electrodynamic process rooted on the IFE but the gyration coefficient can be approximated in the low frequency regime, i.e. $\omega\ll\Gamma$, in terms of the linear complex conductivity, $\sigma_1-i\sigma_2$ based on London equations  \cite{tinkham2004introduction}. 
Note that the magnetic field associated with the microwave field has no contribution to DC magnetization ruling out the direct magnetization of the superconductor. In fact, the linear response of the superconductor due to the incident electromagnetic field, i.e. equations (\ref{Efield}) and (\ref{Bfield}), creates a time-dependent supercurrent and normal current according to Maxwell's and London equations, i.e. ${{\bf \tilde J}}={{\bf \tilde J}}_s+{{\bf \tilde J}}_n=(\sigma_1-i\sigma_2){\bf {\tilde{E}}}$. For a circularly polarized electric field there is a time-dependent circular supercurrent having $x$ and $y$ components. In addition,    
the microwave-induced DC magnetization forms a supercurrent density due to nonlinear response based on IFE and is shown by ${\bf J}^{\text {ind}}_s$. The induced supercurrent due to DC magnetization opposes to the incident electric field to maintain the Meissner effect and can be written as
\begin{equation}
    {\bf J}^{\text{ind}}_{s}=\nabla\times {\bf M}_{DC}= M_{DC}\Big(
    {\bf x}\frac{\partial f(x,y)}{\partial y}-{\bf y}\frac{\partial f(x,y)}{\partial x}\Big)
\end{equation}
where we consider the magnetization profile dictated by microwave radiation in the $xy$ plane to be considered by an arbitrary function $f(x,y)$. The induced DC magnetic vector potential, ${\bf A}^{\text{ind}}$, can also be defined as 
\begin{equation}
\nabla\times {\bf A}^{\text{ind}}=\mu_o{\bf M}_{DC}
\end{equation}
At temperatures much lower than the critical temperature, the microwave field propagates in low-loss regime penetrating the superconductor sample and the strength of DC magnetization decreases exponentially. Although the strength of DC magnetization is more pronounced at lower frequency, i.e. ${\bf M}_{DC}\propto\omega^{-3}$, but lower the frequency leads to a larger radiation area in the order of $\lambda^2$, where $\lambda$ is the free-space wavelength of the microwave field. \\
In type I superconductor, a magnetic field is screened until the critical field $H_c$ is reached. In the case of microwave-induced DC magnetization, the critical field $H_c$ is reached where the superposition of DC magnetization and the applied microwave field amplitude $H_o=\displaystyle{\frac{B_o}{\mu_o}}$ adds up to the critical field. Then there is a critical electric field, $E_{c}$, where the superconducting phase is thermodynamically unstable, and that can be written as
\begin{equation}
E_{c}=\frac{
\bigg(\sqrt{1+16\eta^4_o\gamma^2(\omega,T)H^2_c}-1\bigg)^{1/2}}{2\sqrt{2}\gamma(\omega,T)\eta_o}
\end{equation}
where $\eta_o$ is the free space characteristic impedance.\\

\begin{figure}
\includegraphics[width=8cm,height=5cm]{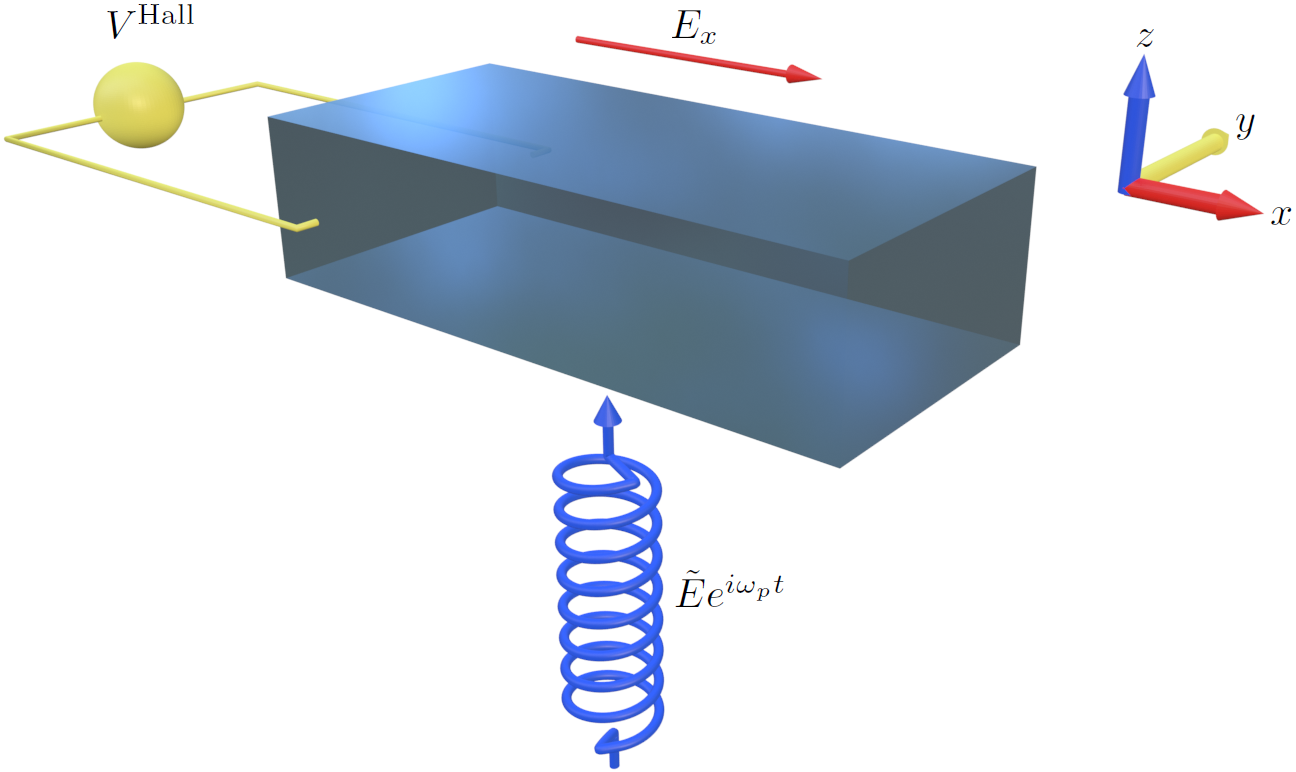}
\caption{ Superconducting wire is illuminated by the circularly polarized microwave pump signal in the presence of DC electric field, $E_x$ develops Hall voltage $V^{\text{Hall}}$.}
\label{Hallfig}
\end{figure}

\noindent
{\it Gyroelectric Conductivity and Hall Effect}-
The microwave-induced DC magnetization breaks the directional symmetry making a superconductor a  gyrotropic material represented by a conductivity tensor, the so-called gyroelectric conductivity. Referring to the London model, one can can find the gyroelectric conductivity tensor, $\bar{\bar\sigma}$, relating the total current density $\bf J$ to the applied weak electric field $\bf E$, i.e. $\bf J=\bar{\bar\sigma}\bf E$. Considering ${\bf \tilde E}={\bf x}\tilde E_x+{\bf y}\tilde E_y+{\bf z}\tilde E_z{\bf z}$, with angular frequency $\omega$, the gyroelectric conductivity tensor can be written as
\begin{equation}
\begin{aligned}
\label{SigmaTensor}
\bar{\bar{\sigma}}(\omega_s)=\left(\begin{array}{ccc} \sigma_s+\sigma_n & -i\displaystyle{\frac{\omega_c}{\omega}}\sigma_s+\tau\omega_c\sigma_n & 0\\ 
i\displaystyle{\frac{\omega_c}{\omega}}\sigma_s-\tau\omega_c\sigma_n & \sigma_s+\sigma_n & 0\\
0 & 0 & \sigma_1-i\sigma_2 \end{array}\right)
\end{aligned}
\end{equation}
where
\begin{eqnarray}
\sigma_s&=&\frac{i\omega e^2n_s}{m(\omega^2_c-\omega^2)}\\
\sigma_n&=&\frac{n_n e^2\tau(1+i\omega\tau)}{m[(1+i\omega\tau)^2+\omega^2_c\tau^2]}
\end{eqnarray}
and the microwave-induced cyclotron angular frequency is \begin{equation}
    \omega_c\triangleq\frac{2e\mu_o}{m}\gamma(\omega_p,T)|E_o|^2
\end{equation}
Note that the cyclotron frequency should be smaller than the gap frequency, otherwise the energy of the microwave-induced magnetic moments is larger than cooper pair binding energy leading to suppressed superconductivity. Therefore, IFE is pronounced in the regime where the microwave pump and signal frequencies are smaller than the cyclotron and gap frequencies and the field amplitudes are smaller than the critical electric field.      
The gyroelectric conductivity can be tracked down due to the contribution of microwave-induced magnetization to the longitudinal conductivities,  $\sigma_{xx}=\sigma_{yy}=\sigma_s+\sigma_n$ through the cyclotron frequency $\omega_c$. Gyroelectric conductivity also leads to the Hall effect where the microwave-induced magnetization causes a DC electric field to develop across the superconducting wire perpendicular to the direction of microwave propagation, i.e. $z$ direction \cite{band2006light}. Consider a long superconducting wire along the $x$ direction with the corresponding length scales at $l_x$, $l_y$ and $l_z$, and in the presence of  longitudinal DC electric field along $x$  direction and a circularly polarized microwave signal propagating along z direction as shown in Figure \ref{Hallfig}. The microwave-induced DC magnetization produces the Hall field along the $y$ direction, where the current cannot flow out of the wire along the y-direction, i.e. $J_y=0$. The Hall resistance is then given by
\begin{equation}
R_{Hall}=\frac{1}{l_z}\frac{\displaystyle{\frac{i\omega_c}{\omega}}\sigma_s-\tau\omega_c\sigma_n}{(\sigma_s+\sigma_n)^2+(\displaystyle{\frac{i\omega_c}{\omega}}\sigma_s-\tau\omega_c\sigma_n)^2}
\Bigg\rvert_{\omega=0}    
\end{equation}
where the microwave field decay has not been considered therefore, the length should be chosen less than the inverse of the field attenuation constant, i.e. $l_z<\alpha^{-1}_p$. The Hall resistance shows the contribution from both superelectrons and normal-electrons in the superconducting wires and well into the superconducting state the Hall resistance will be
\begin{equation}
\label{Hall_s}
R_{Hall}=\frac{\mu_o\lambda^2_L}{l_z}\omega_c    
\end{equation}
where $\lambda_L$ is the London penetration depth. Eq. (\ref{Hall_s}) offers a way for the measurement of the gyration coefficient of the superconductor through the Hall resistance measurement.
Close to the critical temperature the Hall resistance tends to its normal value of 
$R_{Hall}=\displaystyle{\frac{\mu_oM_{DC}}{n_nel_z}=\frac{2\mu_o\gamma(\omega_p,T)}{n_n e l_z}|E_o|^2}$.
This result conforms the linear dependency of the Hall resistance to the microwave-induced DC magnetization while the longitudinal resistance is unaffected by the microwave signal.
\\
{\it Gyroelectric Birefringence}-
Another feature of the gyroelectric conductivity in superconductor is the microwave birefringence. According to Maxwell's equations, the conductivity tensor, Eq. (\ref{SigmaTensor}), yields the anisotropic permittivity tensor as
\begin{equation}
\begin{aligned}
\label{EpsTensor}
\bar{\bar\epsilon}_r=1-\displaystyle{\frac{i}{\omega\epsilon_o}}\bar{\bar\sigma}=\left(\begin{array}{ccc} \epsilon'_{r}-i\epsilon''_{r} & \epsilon'_{xy}-i\epsilon''_{xy} & 0\\ 
-\epsilon'_{xy}+i\epsilon''_{xy} & \epsilon'_{r}-i\epsilon''_{r} & 0\\
0 & 0 & \epsilon'_{zz}-i\epsilon''_{zz} \end{array}\right)
\end{aligned}
\end{equation}
Note that the superconductor's relative permittivity consists of a large negative real part where the electromagnetic field is weak with frequency significantly smaller than its gap frequency and below its critical temperature.     
Now, if a weak linearly-polarized wave with a frequency of $\omega<\omega_c$ is lunched to the superconductor along the $z$ axis copropagating with the microwave pump field with frequency $\omega_p$, it experiences birefringence leading to polarization rotation. If the permittivity tensor is diagonalized in the coordinate system with orthogonal unit vectors ${\bf e}_{\pm}=\displaystyle{\frac{1}{\sqrt{2}}({\bf x}\pm i{\bf y})}$, then the linearly-polarized wave has two normal propagation modes with relative permittivities $(\epsilon'_{xx}\pm\epsilon''_{xy})- i(\epsilon''_{xx}\mp \epsilon'_{xy})$ while
$z$ axis acts as the uniaxial optical symmetry.
The weak signal entering the gyrotropic superconductor decomposes in to the slow component with the complex propagation constants of 
\begin{equation}
\label{fastwave}
\alpha_1+i\beta_1=\frac{k_o}{2}\frac{\epsilon''_{xx}-\epsilon'_{xy}}{\sqrt{\epsilon'_{xx}+\epsilon''_{xy}}}+ik_o\sqrt{\epsilon'_{xx}+\epsilon''_{xy}}
\end{equation}
and the fast component having the following complex propagation constant of 
\begin{equation}
\label{slowwave}
\alpha_2+i\beta_2=k_o\sqrt{\epsilon''_{xy}-\epsilon'_{xx}}+i\frac{k_o}{2}\frac{\epsilon''_{xx}+\epsilon'_{xy}}{\sqrt{\epsilon''_{xy}-\epsilon'_{xx}}}
\end{equation}
where $k_o=\frac{2\pi}{\lambda_o}=\frac{\omega}{c}$ is the free space wavenumber in terms of wavelength $\lambda_o$ and $c$ as the speed of light. 
The dependence of the real and imaginary parts of the relative permittivity on the microwave induced magnetization leads to linear birefringence (Cotton-Mouton effect) and a circular birefringence (Faraday effect), respectively  \cite{shen1984principles}.
In the case of $\omega<\omega_c<\omega_s$ one can find the rotation angle of the weak signal per length, commonly known as the rotatory power, i.e. $\phi$, as
\begin{equation}
\label{rot}
\phi=\frac{|\beta_1-\beta_2|}{2}\approx \frac{\omega_s}{\sqrt{\omega\omega_c}} 
\end{equation}\\
where the approximation is valid for $T<T_c$. Eq. (\ref{rot}) highlights radically different response of superelecrons to the microwave-induced magnetization than normal electrons. For dielectrics and metals, the rotation angle is quadratically dependent on the pump electromagnetic field, i.e. optical field, leading to a Verdet constant \cite{battiato2014quantum,majedi2020nonlinear} that is not the case for superconductors.\\
{\it Flux Quantization}- Consider the macroscopic quantum model of superconductivity where the local density of superelectrons in its ground state can be described by a macroscopic wavefunction $\Psi({\bf r},t)=\sqrt{n_s({\bf r},t)}e^{i\theta({\bf r},t)}$ obeying the Schr\"odinger equation \cite{orlando1991foundation}. In the presence of a magnetic vector potential
the supercurrent reads $\Lambda{\bf J}_{s}({\bf r},t)=\displaystyle{\Big(\frac{\hbar}{2e}\nabla\theta-{\bf A}({\bf r},t)\Big)}$ where $\Lambda\triangleq\displaystyle{\frac{m}{2n_se^2}}$ is the London parameter. 
Suppose we integrate the supercurrent equation about a closed contour within the superconductor where the path is either in bulk superconducting region or multiply connected region in the presence of circularly polarized field. Assuming that $|\Psi({\bf r},t)|$ is a well-defined function then the line integration of supercurrent equation along the contour $C$ encircling the surface $S$ yields the following fluxoid quantization expression
\begin{eqnarray}
\nonumber
\label{Flux_Q}
    \oint_C i\Lambda\gamma(\omega,T) \nabla\times\Big(f(x,y){\bf{\tilde E}}\times{\bf{\tilde E}}^*\Big).d{\bf l}&+&\\
    \mu_o\int_Si\gamma(\omega,T)f(x,y){\bf{\tilde E}}\times{\bf{\tilde E}}^*.d{\bf S}&=&n\Phi_o
\end{eqnarray}
where $\Phi_o$ is the flux quantum, $n$ represents a winding number of the macroscopic wavefunction. The left-hand side of equation (\ref{Flux_Q}) represents the electric field-induced fluxoid in the superconductor. Thus, if the superconductor is magnetized by the circularly polarized microwave field, once the field is removed the trapped flux inside the superconductor is quantized. This might offer new way to magnetize/demagnetize superconductors with incident microwave field.\\
{\it Vortex State}- Consider a type II superconductor where the GL parameter $\kappa=\displaystyle{\frac{\lambda}{\xi}}>\frac{1}{\sqrt{2}}$, that is defined as the ratio of its penetration depth $\lambda$ to its coherence length $\xi$. In the mixed state, the magnetic flux penetrates type II superconductor starting at lower critical field $H_{c1}=\displaystyle{\frac{\Phi_o}{4\pi\mu_o\lambda^2}}\ln(\frac{\lambda}{\xi})$ in the form of triangular array of vortices until reaches its upper critical field $H_{c2}=\displaystyle{\frac{\Phi_o}{2\pi\mu_o\xi^2}}$. In order to derive the type II superconductor to its mixed states by the microwave-induced IFE, we need to satisfy the inequality relation for the electric field amplitude as $E_{c1}<E_o<E_{c2}$ where
\begin{equation}
E_{c1,2}=
   \frac{\bigg(\sqrt{1+16\eta^4_o\gamma^2(\omega,T)H^2_{c1,2}}-1\bigg)^{1/2}}
   {2\sqrt{2}\gamma(\omega,T)\eta_o}
\end{equation}
By turning on/off the microwave field and controlling the pump frequency and its amplitude one can dynamically generate and tune vortices in the superconductors without applying magnetic field.\\ 
For Nb with $\xi=38$nm and $\lambda_L=39$nm at zero temperature the corresponding critical magnetic fields are $B_{c1}=2.8$ mT,  $B_{c}=200$ mT and $B_{c2}=228$ mT, leading to the critical electric fields $E_{c1}=28.55$ $\mu$V/m, $E_c=240$ $\mu$V/m  and $E_{c2}=257$ $\mu$V/m at 3 Mrad/s pump angular frequency. The cyclotron frequency for the first critical field in Nb is 0.49 Grad/s. This example reveals that microwave-induced IFE can happen at low-power electric field excitation in absence of any other stimulation such as biasing current or magnetic field.\\
{\it Conclusion}- I have explored the possibility of IFE in superconductors as a new source of nonlinear and gyrotropy. IFE and its associated phenomena provide new avenues for microwave-induced non equilibrium and chiral states potentially useful for superconducting microwave devices, e.g. tunable resonators and detectors. They offer novel applications to control superconductivity in a dynamic and  fully controllable fashion that is solely enabled by a microwave polarization degree of freedom. New readout electronics for superconducting qubits and cavity QED circuits can be envisioned where a train of linearly and circularly polarized microwave pulses can potentially control the timing of qubit initialization, computation and measurement.   
\nocite{*}

\bibliography{apssamp}

\end{document}